\documentclass[a4paper,dvips,12pt]{article}
\usepackage{amsmath,latexsym,amssymb,epsfig,psfrag}  
\usepackage{comment}
\usepackage{cite}
\usepackage{wasysym}

\def\ltsim{\lower3pt\hbox{$\, \buildrel < \over \sim \, $}}  
\def\gtsim{\lower3pt\hbox{$\, \buildrel > \over \sim \, $}}  
\newcommand{\be}{\begin{equation}}  
\newcommand{\ee}{\end{equation}}  
\def\ga{\mathrel{\raise.3ex\hbox{$>$\kern-.75em\lower1ex\hbox{$\sim$}}}}  
\def\la{\mathrel{\raise.3ex\hbox{$<$\kern-.75em\lower1ex\hbox{$\sim$}}}}

\makeatletter   
\@addtoreset{equation}{section}   
\makeatother

\def\H{{\cal H}}

\def\S{{\cal S}}
\def\be{\begin{equation}}
\def\ee{\end{equation}}
\def\bea{\begin{eqnarray}}
\def\eea{\end{eqnarray}}

\def\deu{{\delta}^{(1)}}
\def\ded{{\delta}^{(2)}}

\begin{document}  
  
\baselineskip=16pt   
\begin{titlepage}  
\begin{center}  
\hfill{{ DFPD-A-03-32}}

\vspace{0.5cm}  
  
\Large {\bf \Large Enhancement of Non-Gaussianity after Inflation}
  
\vspace{2cm}  
\normalsize  
  
{\sc \large N. Bartolo$^{a,}$\footnote{nb21@pact.cpes.susx.ac.uk}, 
S. Matarrese$^{b,}$\footnote{sabino.matarrese@pd.infn.it} and  
A.~Riotto$^{b,}$\footnote{antonio.riotto@pd.infn.it}}   

\smallskip   
\medskip   
\it{~$^{a}$~Astronomy Centre, University of Sussex}\\ 
\it{Falmer, Brighton, BN1 9QJ, U.K.}

\smallskip    
\medskip  
\it{~$^{b}$~Department of Physics and INFN,\\
Sezione di Padova, via Marzolo 8,
I-35131 Padova, Italy}

\vskip0.6in \end{center}  
   
\centerline{\large\bf Abstract}  
\vskip 1cm

\noindent
We study the evolution of cosmological  perturbations on large scales,
up to second
order, for a perfect fluid with generic equation of state. Taking advantage
of super-horizon conservation laws, 
it is possible to 
 follow the evolution of the non-Gaussianity of  perturbations
through the different stages after inflation.
We find that a large non-linearity is generated by the 
gravitational dynamics from the original inflationary quantum
fluctuations. This leads to 
a significant enhancement 
of the tiny  intrinsic 
non-Gaussianity   produced during inflation  in single-field 
slow-roll models.

\vspace*{2mm}

\end{titlepage}  
  
\section{\sc Introduction}  \label{sec:intro}

Inflation is the simplest and most successful mechanism proposed to date 
for the causal generation of primordial cosmological perturbations 
on  cosmologically relevant scales \cite{lr}. 
The gravitational amplification of the primordial perturbations is supposed
to  seed structure formation in the Universe and produce 
Cosmic Microwave Background (CMB) anisotropies in  agreement 
with observational data \cite{wmap}.  
Due to  the smallness of the primordial cosmological perturbations,  
their generation and evolution have usually been studied within linear theory
\cite{revp}.
After the seminal work by Tomita \cite{tomita}, 
only recently second-order perturbation theory \cite{BMMS,MMB,noh}
has been employed
to evaluate specific physical observables generated during
inflation \cite{noi,maldacena}.  The importance of an accurate 
determination 
of higher-order statistics as the bispectrum comes from the 
fact that they allow   to search for 
the signature of non-Gaussianity in the
primordial perturbations which is usually parametrized by
a dimensionless non-linear parameter $f_{\rm NL}$. 
Indeed, 
a number of present and future CMB experiments, such as 
{\it WMAP} \cite{k}
 and {\it Planck}, have enough resolution to either constrain or detect 
non-Gaussianity of CMB anisotropy data with high precision \cite{ks}.

The main result of the second-order 
analysis performed in \cite{noi,maldacena} is that
single-field slow-roll models of inflation give rise to a 
level of intrinsic non-Gaussianity which -- at the end of the inflationary 
stage -- is tiny,  being first-order in the slow-roll parameters. 

The goal of this paper is to 
study the post-inflationary evolution on super-horizon scales 
of the primordial 
non-linearity in the cosmological  perturbations. We perform a
fully relativistic analysis of the dynamics
of second-order perturbations for a perfect fluid with  generic 
equation of state taking advantage of the super-horizon
conservation of the second-order gauge-invariant curvature 
perturbation recently discussed in Refs. \cite{lw,mw} (see also
\cite{noi,nak,rigo}).
Our main result is that the post-inflationary evolution
gives rise to an enhancement of the level of non-Gaussianity on super-horizon
scales.  
Once again, inflation  provides the key generating mechanism to 
produce super-horizon seeds, which are later amplified by gravity. 

The plan of the paper is as follows. In Section 2 we provide the second-order
expansion of the metric and of the energy-momentum tensor, assuming that
the source term is represented by a perfect fluid with constant equation of 
state. In Section 3 we solve the perturbed Einstein equations up to 
first order around a Friedmann-Robertson-Walker background. 
The body of the paper is contained in 
Section 4, where we derive the super-horizon evolution equations
of the second-order gravitational potential and density perturbations.
Section 5 contains a brief discussion of our findings.

\section{\sc 
Perturbations of a flat Robertson-Walker Universe up to second order}
In order to study the perturbed Einstein equations, 
we first write down the perturbations on a spatially flat 
Robertson-Walker background following the formalism of Refs.\cite{BMMS,MMB}. 
We shall first consider the fluctuations of the metric, and then the 
fluctuations of the energy-momentum tensor. Hereafter 
greek indices run from $0$ to $3$, while latin indices label  
the spatial coordinates from $1$ to $3$. If not otherwise specified we 
will work with conformal time $\tau$, and a prime will stand for a derivative
with respect to $\tau$.

\subsection{The metric tensor}
The components of a perturbed spatially flat Robertson-Walker 
metric can be written as
\bea \label{metric1}
g_{00}&=&-a^2(\tau)\left( 1+2 \phi^{(1)}+\phi^{(2)} \right)\, ,\nonumber\\
g_{0i}&=&a^2(\tau)\left( \hat{\omega}_i^{(1)}+\frac{1}{2} 
\hat{\omega}_i^{(2)} \right)
\, ,  \nonumber\\g_{ij}&=&a^2(\tau)\left[
(1 -2 \psi^{(1)} - \psi^{(2)})\delta_{ij}+
\left( \hat{\chi}^{(1)}_{ij}+\frac{1}{2}\hat{\chi}^{(2)}_{ij} \right)\right] 
\,,
\eea
where the scale factor $a$ is a function of the conformal time
$\tau$. 
The standard splitting of the perturbations into scalar, transverse 
({\it i.e} divergence-free) vector parts, and transverse trace-free tensor 
parts with respect to the 3-dimensional space with metric $\delta_{ij}$ 
can be performed in the following way:
\begin{equation}
\hat{\omega}_i^{(r)}=\partial_i\omega^{(r)}+\omega_i^{(r)}\, ,
\end{equation}
\begin{equation}
\hat{\chi}^{(r)}_{ij}=D_{ij}\chi^{(r)}+\partial_i\chi^{(r)}_j
+\partial_j\chi^{(r)}_i
+\chi^{(r)}_{ij}\, ,
\end{equation}
where $(r)=(1),(2)$ stand for the order of the perturbations, $\omega_i$
and $\chi_i$ are transverse vectors ($\partial^i\omega^{(r)}_i=
\partial^i\chi^{(r)}_i=0$), $\chi^{(r)}_{ij}$ is a symmetric transverse and 
trace-free tensor ($\partial^i\chi^{(r)}_{ij}=0$, $\chi^{i(r)}_{~i}
=0$) and $D_{ij}=\partial_i \partial_j - (1/3) \, \, 
\delta_{ij}\, \partial^k\partial_k$ is a trace-free operator.
Here and in the following latin indices 
are raised and lowered using $\delta^{ij}$ and $\delta_{ij}$, respectively.\\
For our purposes the metric in Eq.~(\ref{metric1}) can be simplified. In fact, 
first-order vector perturbations are zero; 
moreover, the tensor part gives a negligible contribution to second-order
perturbations. Thus, in the following we can neglect  
$\omega^{(1)}_i$, $\chi^{(r)}_{(1)i}$ and $\chi^{(r)}_{(1)ij}$.
However the same is not true for the second order perturbations. 
In the second-order theory the second-order vector and tensor 
contributions can be generated by the first-order scalar perturbations 
even if they are initially zero 
\cite{MMB}. Thus we have to take them into account and we shall use the metric
\bea \label{metric2}
g_{00}&=&-a^2(\tau)\left( 1+2 \phi^{(1)}+\phi^{(2)} \right)\, ,\nonumber\\
g_{0i}&=&a^2(\tau)\left( \partial_i\omega^{(1)}+\frac{1}{2}\, 
\partial_i\omega^{(2)}+\frac{1}{2}\, \omega_i^{(2)} \right)
\, ,  \nonumber\\g_{ij}&=&a^2(\tau)\left[
\left( 1 -2 \psi^{(1)} - \psi^{(2)} \right)\delta_{ij}+
D_{ij}\left( \chi^{(1)} +\frac{1}{2} \chi^{(2)} \right)\right.\nonumber\\
&+&\left.\frac{1}{2}\left( \partial_i\chi^{(2)}_j
+\partial_j\chi^{(2)}_i
+\chi^{(2)}_{ij}\right)\right]
\,.
\eea 
The controvariant metric tensor is obtained by requiring (up to second order)
that $g_{\mu\nu}g^{\nu\lambda}=\delta_\mu\, ^\lambda$ and it is given by
\bea \label{cont}
g^{00}&=&-a^{-2}(\tau)\left( 1-2 \phi^{(1)}-\phi^{(2)} +4\left(
\phi^{(1)}\right)^2-\partial^i\omega^{(1)}\partial_i\omega^{(1)}  
\right)\, ,\nonumber\\
g^{0i}&=&a^{-2}(\tau)\left[ \partial^i\omega^{(1)}+\frac{1}{2}
\left( \partial^i\omega^{(2)}+\omega^{i(2)} \right) +2 \left( \psi^{(1)}
-\phi^{(1)} \right) \partial^i\omega^{(1)}-\partial^i\omega^{(1)}
D^i\,_k \chi^{(1)} \right]
\, ,  \nonumber\\
g^{ij}&=&a^{-2}(\tau) \left[
\left( 1+2 \psi^{(1)} +\psi^{(2)}+4 \left( \psi^{(1)} \right)^2 
\right) \delta^{ij}-
D^{ij}\left( \chi^{(1)} +\frac{1}{2} \chi^{(2)} \right)\right.\nonumber \\
&-& \frac{1}{2}\left( \partial^i\chi^{j(2)}
+\partial^j\chi^{i(2)}
+\chi^{ij(2)} \right)-\partial^i\omega^{(1)}\partial^j\omega^{(1)}\nonumber\\
&-&\left.
4 \psi^{(1)}D^{ij}\chi^{(1)}+D^{ik}\chi^{(1)}D^j_{~k} \chi^{(1)} \right]\,.
\eea
Using $g_{\mu\nu}$ and $g^{\mu\nu}$ one can calculate the 
connection coefficients and the Einstein 
tensor components up to second order in the metric 
fluctuations. They are given in the Appendix A of Ref. \cite{noi}.
From now one, we will adopt the {\it Poisson
gauge} \cite{poisson} which is defined
by the condition $\omega=\chi=\chi_{i}=0$. Then, one scalar
degree of freedom is eliminated from $g_{0i}$ and one scalar
and two vector degrees of freedom from $g_{ij}$. This gauge generalizes the
so-called longitudinal gauge to include vector and tensor modes
and contains a solenoidal vector $\omega_i^{(2)}$.

\subsection{Energy-momentum tensor of the fluid}
Since after inflation and reheating the Universe enters 
a radiation-dominated phase and, subsequently, a matter- 
and dark energy-dominated phases, 
we shall consider a generic fluid characterized by an energy density
$\rho$ and pressure $P$ with energy-momentum tensor

\begin{equation}
\label{EM}
T^\mu_{~ ~\nu}= \left(\rho+P\right)u^\mu u_\nu + P \delta^{\mu}_{~\nu}\,,
\end{equation}
where $u^\mu$ is the four-velocity vector subject to the constraint
$g^{\mu\nu}u_\mu u_\nu=-1$. At second order
of perturbation theory it can be decomposed as

\begin{equation}
u^\mu=\frac{1}{a}\left(\delta^{\mu}_0+v^{\mu}_{(1)}+\frac{1}{2}v^{\mu}_{(2)}
\right)\, .
\end{equation}
For the first- and second-order perturbations, we get 

\bea
v^{0}_{(1)}&=&- \psi^{(1)},\nonumber\\
v^{0}_{(2)}&=& -\phi^{(2)} +3\left(\psi^{(1)}\right)^2+ v_i^{(1)}
v^i_{(1)}\,.
\eea
Similarly, we obtain

\bea
u_{0}&=&a\left(-1-\phi^{(1)}-\frac{1}{2}\phi^{(2)}+\frac{1}{2}
\left(\psi^{(1)}\right)^2-\frac{1}{2}v_i^{(1)}
v^i_{(1)}\right)\,,\nonumber\\
u_{i}&=&a\left(v_i^{(1)}+\frac{1}{2}v_i^{(2)}-2\psi^{(1)}v_i^{(1)}
+\frac{1}{2}\omega_i^{(2)}\right)\, .
\eea
The energy density $\rho$ 
can be split into a homogeneous background $\rho_0(\tau)$ 
and a perturbation $\delta \rho (\tau, x^{i})$ as follows
\begin{equation}
 \label{rho}
\rho(\tau, x^i)=\rho_0(\tau)+\delta\rho(\tau, x^i)=\rho_0(\tau)+
\delta^{(1)}\rho(\tau, x^i)+\frac{1}{2}\delta^{(2)}
\rho(\tau, x^i)\, ,
\end{equation}
where the perturbation has been expanded into a first and a second-order
part, respectively. The same decomposition can be adopted for the
pressure $P$.

Using the expression (\ref{rho}) into Eq. (\ref{EM}) and calculating
$T^\mu_{~\nu}$ up to second order we find 
\bea 
T^\mu_{~\nu}=T^{\mu(0)}_{~\nu}+\deu T^{\mu}_{~\nu}+\ded
T^{\mu}_{~\nu} \, ,
\eea
where $T^{\mu(0)}_{~\nu}$ corresponds to the background value, and
\bea 
\label{scalT_00}
T^{0(0)}_{~0}+\deu T^{0}_{~0}&=& -\rho_0-\delta^{(1)}\rho\, , \\ 
\label{00}
\ded T^0_{~0}&=& -\frac{1}{2}\delta^{(2)}\rho-\left(1+w\right)\rho_0
v_i^{(1)}v^i_{(1)}\, ,  \\
&& \nonumber \\
&& \nonumber \\
\label{scalT_0i}
T^{i(0)}_{~0}+\deu T^{i}_{~0}&=& -\left(1+w\right)\rho_0 v^i_{(1)}\, , \\
\label{0i}
\ded {T^i_{~0}}&=& -\left(1+w\right)\rho_0\left[\left(\psi^{(1)}+
\frac{\delta^{(1)}\rho}{\rho_0}\right)v^i_{(1)}+\frac{1}{2}v^i_{(2)}\right]
\, ,\\
&& \nonumber \\
&& \nonumber \\
T^{i(0)}_{~j}+\deu T^{i}_{~j}&=& w\rho_0 \left(1+
\frac{\delta^{(1)}\rho}{\rho_0}\right)
\delta^{i}_{~j}\,, \\
\label{ij}
\ded{T^i_{~j}}&=& \left(1+w\right)\rho_0v^i_{(1)}v_j^{(1)}+
\frac{1}{2}w\delta^{(2)}\rho\delta^{i}_{~j}\, .
\eea
In the previous expressions we have made the assumption that
the pressure $P$ can be expressed in terms of the energy density as
$P=w\rho$ with constant $w$.

\section{\sc Basic first-order Einstein equations on large-scales}

Our starting point are the 
perturbed Einstein equations $\delta G^\mu_{~\nu}= \kappa^2\, 
\delta T^\mu_{~\nu}$ in the Poisson gauge. Here $\kappa^2\equiv
8\pi \,G_{\rm N}$.
At first-order, the  $(0-0)$- and $(i-0)$-components of Einstein equations
are  

\bea
\frac{1}{a^2}
\Bigg[ 6\,\H^2 \phi^{(1)} \,+6\,\H{\psi^{(1)}}^{\prime} -2
\nabla^2\psi^{(1)}\Bigg]
&=& -\kappa^2\delta^{(1)}\rho\, , \\
\label{i-0}
\frac{2}{a^2}\left(\H \partial^i \phi^{(1)} 
+\partial^i
{\psi^{(1)}}^{\prime}\right)&=&\,-\kappa^2
\left(1+w\right)\rho_0 v^i_{(1)}\, , 
\eea
where we have indicated by $\H=\frac{a^\prime}{a}$ the Hubble rate 
in conformal time. These equations, together 
with the non-diagonal part of the $(i-j)$-component
of Einstein equations, give $\psi^{(1)}=\phi^{(1)}$ and, on 
super-horizon scales,  

\begin{equation}
\label{fo}
\psi^{(1)}=-\frac{1}{2}\frac{\delta^{(1)}\rho}{\rho_0}
=\frac{3(1+w)}{2}\H \frac{\delta^{(1)}\rho}{\rho^\prime_0}
\, .
\end{equation}
The continuity equation yields an evolution equation
for the large-scale energy density perturbation 

\begin{equation}
\label{ll}
\delta^{(1)}\rho^\prime+3\H\left(1+w\right)\delta^{(1)}\rho-3
\psi^{(1)^\prime}\left(1+w\right)\rho_0 =\frac{2}{3}\frac{\rho_0}{\H^2}
\nabla^2\left(\psi^{(1)\prime}+\H\psi^{(1)}\right)
\, .
\end{equation}
This equation, together with the the background continuity equation
$\rho_0^\prime+3\H\left(1+w\right)\rho_0=0$, leads to the
conservation on large-scales of the first-order gauge-invariant
curvature perturbation \cite{revp}

\begin{equation}
\zeta^{(1)}=-\psi^{(1)}-\H\frac{\delta^{(1)}\rho}{\rho^\prime_0}\, .
\end{equation}
Indeed, both the density perturbation, $\delta\rho$ and the
curvature perturbation, $\psi$, are in general gauge-dependent.   
Specifically,  they depend upon the chosen time-slicing in an
inhomogeneous universe. The curvature perturbation on fixed time hypersurfaces
is a gauge-dependent quantity: after an arbitrary linear coordinate
transformation at first-order, $t\rightarrow t+\delta t$, it transforms as
$\psi^{(1)}\rightarrow \psi^{(1)}+\H\delta t$. 
For a scalar quantity, such as the
energy density, the corresponding transformation is $\delta\rho^{(1)}
\rightarrow
\delta\rho^{(1)}-\rho_0^\prime\delta t$.
However the  gauge-invariant combination $\zeta^{(1)}$ can   
be constructed which describes the density perturbation on uniform
curvature slices or, equivalently the curvature of uniform density 
hypersurfaces. 
On large scales $\zeta^{(1)\prime}\simeq 0$. Using Eq. (\ref{fo})
and the background continuity equation, 
we can determine

\begin{equation}
\psi^{(1)}=-\frac{3(1+w)}{5+3 w}\, \zeta^{(1)}\, ,
\end{equation}
which is useful to relate the curvature $\psi^{(1)}$ 
during either the matter 
or the radiation epoch to the gauge-invariant curvature
perturbation $\zeta^{(1)}$ at the end of 
 the inflationary stage. Indeed, since
$\zeta^{(1)}$ is constant,  we can write

\begin{equation}
\label{p}
\psi^{(1)}=-\frac{3(1+w)}{5+3 w}\, \zeta_I^{(1)}\, ,
\end{equation}
where
the subscript ``$I$" means that $\zeta^{(1)}$ is computed during
the inflationary stage.

\section{\sc Basic second-order Einstein equations on large-scales
and  non-Gaussianity}

In order to determine the non-Gaussianity of the cosmological
perturbations after inflation, 
we have to derive the behaviour on large-scales
of the metric  and the energy density perturbations at second order. 
Again, our starting point are the 
Einstein equations perturbed at second order $\delta^{(2)}G^\mu_{~\nu}=
\kappa^2 \, \delta^{(2)}T^\mu_{~\nu}$ in the  
Poisson gauge. The second-order expression
for the Einstein tensor $\delta^{(2)}G^\mu_{~\nu}$ can
be found in any gauge in the Appendix A of Ref. \cite{noi} and we do not
report it here. 

\begin{itemize}

\item The  $(0-0)$-component of Einstein equations
(see Eq. (A.39) in Ref. \cite{noi})
leads to

\bea
\label{z}
&&3\H^2\phi^{(2)}+3\H\psi^{(2)\prime}-\nabla^2\psi^{(2)}-12\H^2
\left(\psi^{(1)}\right)^2-3\left(\nabla\psi^{(1)}\right)^2\nonumber\\
&&-8\psi^{(1)}\nabla^2\psi^{(1)}-3\left(\psi^{(1)\prime}\right)^2=
\kappa^2a^2\ded T^0_{~0}\, .
\eea

\item  A relation between the gravitational potentials at second-order
$\psi^{(2)}$ and $\phi^{(2)}$ can be obtained from the traceless
part of the $(i$-$j)$ components of Einstain's equations 
(see  
Eqs. (A.42) and (A.43) in Ref. \cite{noi}). 
We find

\begin{eqnarray}
\label{constraint}
\psi^{(2)} - \phi^{(2)} & = & - 4 \left(\psi^{(1)}\right)^2 
- \nabla^{-2} \left(2 \partial^i \psi^{(1)} \partial_i
\psi^{(1)} + 3 \left(1+w\right) {\cal H}^2 v_{(1)}^i 
v_{(1)i} \right) \nonumber \\
& + & 3 \nabla^{-4} \partial_i \partial^j 
\left(2 \partial^i \psi^{(1)} \partial_j
\psi^{(1)} + 3 \left(1+w\right) {\cal H}^2 v_{(1)}^i 
v_{(1)j} \right) \; .
\end{eqnarray}
This constraint is the second-order equivalent of the linear constraint
$\psi^{(1)}=\phi^{(1)}$ in the Poisson gauge.


\item In order to close the system and fully determine the variables
$\psi^{(2)}$, $\phi^{(2)}$ and $\delta^{(2)}\rho$, we use the energy
conservation at second-order and the divergence of the $(i-0)$-component
of Einstein equations (see Eq. (A.40) in Ref. \cite{noi})\footnote{Notice
that Eq. (\ref{conservation}) generalizes  Eq. (5.33) of Ref. 
\cite{mw} and corrects a sign misprint in front of the fourth term of 
that equation.}

\bea
\label{conservation}
\delta^{(2)}\rho^\prime&+&3\H\left(1+w\right)\delta^{(2)}\rho
-3\left(1+w\right)\rho_0\psi^{(2)^\prime}-
6(1+w)\psi^{(1)^\prime}\left[\delta^{(1)}\rho
+2\rho_0\psi^{(1)}\right]\nonumber\\
&=&-2(1+w)\rho_0\left(v_i^{(1)}v^i_{(1)}\right)^\prime-
2(1+w)(1-3w)\H\rho_0 v_i^{(1)}v^i_{(1)}\nonumber\\
&+&4(1+w)\rho_0\partial_i\psi^{(1)}v^i_{(1)}+
2\frac{\rho_0}{\H^2}\left(\psi^{(1)}\nabla^2\psi^{(1)\prime}-
\psi^{(1)\prime}\nabla^2\psi^{(1)}\right)\, .
\eea
This equation can be rewritten in a more suitable form
\bea
\label{qq}
&&\left[\psi^{(2)}+\H\frac{\delta^{(2)}\rho}{\rho_0^\prime}
+(1+3w)\H^2\left(\frac{\delta\rho^{(1)}}{\rho_0^\prime}\right)^2-4
\H\left(\frac{\delta\rho^{(1)}}{\rho_0^\prime}\right)\psi^{(1)}
\right]^\prime = 
\frac{2}{3}\left(v_i^{(1)}v^i_{(1)}\right)^\prime 
\nonumber\\
&& + \frac{2}{3}(1-3w)\H
v_i^{(1)}v^i_{(1)} -\frac{4}{3}\partial_i\psi^{(1)}v^i_{(1)} + 
\frac{16}{27\left(1+w\right)^2\H}  \psi^{(1)} \nabla^2 \psi^{(1)}
\nonumber\\
&&
- \frac{2}{3\left(1+w\right)\H^2}\left[\left(1- 
\frac{8}{9\left(1+w\right)}\right) 
\psi^{(1)} \nabla^2\psi^{(1)\prime}
- \left(1 - \frac{4\left(1+3w\right)}{9\left(1+w\right)}\right)
\psi^{(1)\prime} \nabla^2\psi^{(1)} \right] 
\nonumber\\
&&
+ \frac{8\left(1+3w\right)}{27\left(1+w\right)^2 \H^3} 
\left[\frac{\left(\nabla^2 \psi^{(1)}\right)^2}{3}
- \psi^{(1)\prime} \nabla^2 \psi^{(1)\prime} 
+ \frac{\nabla^2 \psi^{(1)\prime} \nabla^2 \psi^{(1)}}{3\H}
\right] \, ,
\eea
where the argument on the left-hand side can be further simplified to 
\bea
\label{3}
&&\psi^{(2)}+\H\frac{\delta^{(2)}\rho}{\rho_0^\prime}
-\left(5+3w\right)\H^2 \left(\frac{\delta^{(1)}\rho}{\rho_0^\prime}
\right)^2\, \nonumber\\
&&=\psi^{(2)}+\H\frac{\delta^{(2)}\rho}{\rho_0^\prime}
-\frac{4}{5+3w}\left(\zeta_I^{(1)}\right)^2\, 
\eea
and the final form has been obtained 
employing Eq. (\ref{p}) \cite{mw1}.
From Eqs. (\ref{qq}) and (\ref{3}) we find

\begin{equation}
\label{three}
\psi^{(2)}+\H\frac{\delta^{(2)}\rho}{\rho_0^\prime}
-\left(5+3w\right)\H^2 \left(\frac{\delta^{(1)}\rho}{\rho_0^\prime}
\right)^2={\cal C} +\frac{2}{3} \left(v_i^{(1)}v^i_{(1)}\right)
+\int^\tau \,d\tau^\prime\, {\cal S}(\tau^\prime)\, , 
\end{equation}
where ${\cal C}$ is a constant in time, 
${\cal C}^\prime=0$, on large-scales, and 

\bea
{\cal S}&=&
\frac{2}{3}(1-3w)\H
v_i^{(1)}v^i_{(1)} -\frac{4}{3}\partial_i\psi^{(1)}v^i_{(1)} + 
\frac{16}{27\left(1+w\right)^2\H}  \psi^{(1)} \nabla^2 \psi^{(1)}
\nonumber\\
&-&
\frac{2}{3\left(1+w\right)\H^2}\left[\left(1- 
\frac{8}{9\left(1+w\right)}\right) 
\psi^{(1)} \nabla^2\psi^{(1)\prime}
- \left(1 - \frac{4\left(1+3w\right)}{9\left(1+w\right)}\right)
\psi^{(1)\prime} \nabla^2\psi^{(1)} \right] 
\nonumber\\
&+&
\frac{8\left(1+3w\right)}{27\left(1+w\right)^2 \H^3} 
\left[\frac{\left(\nabla^2 \psi^{(1)}\right)^2}{3}
- \psi^{(1)\prime} \nabla^2 \psi^{(1)\prime} 
+ \frac{\nabla^2 \psi^{(1)\prime} \nabla^2 \psi^{(1)}}{3\H}
\right] \, .
\eea
\end{itemize}

\subsection{Determination of the non-linearity parameter}
Since we are interested in the determination  of the non-linear parameter
$f^\phi_{\rm NL}$ after the inflationary stage, it is convenient to
fix the
constant  ${\cal C}$ by matching the conserved
quantity at the end of
inflation ($\tau=\tau_I$)
\begin{equation}
\label{four}
{\cal C}= \psi_I^{(2)}
+\H_I\frac{\delta^{(2)}\rho_I}{\rho_{0I}^\prime}-
2\left(\zeta_I^{(1)}\right)^2\, ,
\end{equation}
where we have used the fact that during inflation $w_I\simeq -1$. 

The inflationary quantity  $\left(\psi_I^{(2)}
+\H_I\frac{\delta^{(2)}\rho_I}{\rho_{0I}^\prime}\right)$ 
has been computed in Refs. \cite{noi,maldacena} 
\begin{equation}
\psi_I^{(2)} +\H_I\frac{\delta^{(2)}\rho_I}{\rho_{0I}^\prime}
\simeq \left( \eta - 3 \epsilon \right) \left( \zeta_I^{(1)} \right)^2  
+ {\cal O}(\epsilon,\eta)\, \left({\rm non-local}\,\,{\rm terms}\right)
\, ,
\end{equation}
in terms of the slow-roll parameters $\epsilon=1-\H_I^\prime/\H_I^2$
and 
$\eta = 1+\epsilon -\left(\varphi^{\prime\prime}/\H_I\varphi^\prime\right)$ 
where $\H_I$ is the Hubble parameter during inflation and 
$\varphi$ is the inflaton field driving the exponential
growth of the scale factor during inflation \cite{lr}. 
Since during inflation the slow-roll parameters are tiny, we can safely
disregard the intrinsically second-order 
terms originated from the inflationary epoch.

Combining Eqs.~(\ref{z}),~(\ref{constraint}),~(\ref{three}) and~(\ref{four})
we single out an equation for the gravitational potential $\phi^{(2)}$ 
on large scales
\bea
\label{new}
\phi^{(2)\prime}+\frac{5+3 w}{2} \H \phi^{(2)} &=& (5+3 w) \H  \left( 
\psi^{(1)} \right)^2 +\frac{3}{2} \H (1+w) 
\left[
\nabla^{-2} \left(2 \partial^i \psi^{(1)} \partial_i
\psi^{(1)} \right. \right.\nonumber \\ 
&+& \left. \left. 3 \left(1+w\right) {\cal H}^2 v_{(1)}^i 
v_{(1)i} \right)
- 3 \nabla^{-4} \partial_i \partial^j 
\left(2 \partial^i \psi^{(1)} \partial_j
\psi^{(1)} \right. \right.\nonumber \\ 
&+& 3 \left. \left. \left(1+w\right) {\cal H}^2 v_{(1)}^i 
v_{(1)j} \right)\right]    
+ \frac{3}{2} \H (1+w) \int^{\tau}_{\tau_I} \S(\tau') d\tau'+\frac{1}{\H} 
\left( \nabla\psi^{(1)} \right)^2\nonumber \\ 
&+& \frac{8}{3\H} \psi^{(1)} \left( \nabla^2 \psi^{(1)} \right) 
+ \frac{\nabla ^2 \S_1}{3\H}+ 
\frac{1}{\H} \left( \psi^{(1)\prime} \right)^2 -\S_1^{\prime} \, ,
\nonumber \\
\eea
where $\S_1$ denotes the R.H.S. of Eq~(\ref{constraint}).

We want to integrate this equation from $\tau_I$ to a time $\tau$ in 
the matter-dominated epoch. The general solution is given by the solution of 
the homogeneous equation plus a particular solution
\bea
\phi^{(2)}& = &
\phi^{(2)}(\tau_I) \exp\left[ - \int_{\tau_I}^\tau \frac{5+3 w}{2} \H d\tau'
\right ]\nonumber \\
&+&\exp\left[-\int_{\tau_I}^\tau \frac{5+3 w}{2} \H d\tau'
\right ] \times \int_{\tau_I}^{\tau} 
\exp \left[ \int_{\tau_I}^{\tau'} \frac{5+3 w}{2} \H ds \right]  
b(\tau') d\tau'\, ,
\eea    
where $b(\tau)$ stands for the source term in the R.H.S of Eq.~(\ref{new}).
Notice that the homogeneous solution during both the radiation and the 
matter-dominated epoch decreases in time. 
Therefore we can neglect the homogeneous solution and 
focus on the contributions from the source term $b(\tau)$.
At a time $\tau$ in the matter-dominated epoch 
$ \exp [ - \int_{\tau_I}^\tau d\tau'\, \H\, (5+3 w)/2] 
\propto \tau^{-5}$.
Thus if we are interested in the 
gravitational potential 
$\phi^{(2)}$ during the matter dominated epoch the contributions in the 
particular solution coming from the 
radiation-dominated epoch can be considered negligible.   
Recalling that during the matter-dominated epoch the linear gravitational 
potential $\psi^{(1)}$ is constant in time, it turns out that 
\bea
\label{expr:phi2}
\phi^{(2)}& \simeq & 2 \left( \psi^{(1)} \right)^2+\frac{3}{5} 
\left[ \nabla^{-2} \left( \frac{10}{3} \partial^i \psi^{(1)} \partial_i
\psi^{(1)} \right) 
-3 \nabla^{-4} \partial_i \partial^j 
\left(\frac{10}{3}  \partial^i \psi^{(1)} \partial_j
\psi^{(1)} \right)\right] \nonumber \\
&+& \exp\left[-\int_{\tau_I}^\tau \frac{5+3 w}{2} \H d\tau'
\right ] \times \int_{\tau_I}^{\tau} 
\exp \left[ \int_{\tau_I}^{\tau'} \frac{5+3 w}{2} \H ds \right]  
 \left \{
\frac{3}{2} \H (1+w) \int^{\tau'}_{\tau_I} \S(s) ds \nonumber \right. \\
&+& \left. \frac{1}{\H} 
\left( \nabla\psi^{(1)} \right)^2 
+\frac{8}{3\H} \psi^{(1)} \left( \nabla^2 \psi^{(1)} \right)
+\frac{\nabla^2 \S_1}{3 \H}  
\right \} d \tau'\, ,
\eea 
where we have used Eq.~(\ref{i-0}) to express the first-order velocities 
in terms of the gravitational potential, and we have taken into account 
that during the matter-dominated epoch $\S_1'=0$. 

As the gravitational potential $\psi^{(1)}$ on super-horizon scales is
generated during inflation, it is clear that the origin of the non-linearity
traces back to the inflationary quantum fluctuations.

The total curvature perturbation will then have a
non-Gaussian
$(\chi^2)$-component. For instance, the lapse function 
$\phi=\phi^{(1)}+\frac{1}{2}\phi^{(2)}$ can be expressed in momentum space
as 
\begin{equation}
\label{phimomspace}
\phi({\bf k}) = \phi^{(1)}({\bf k}) + 
\frac{1}{(2\pi)^3}
\int\, d^3 k_1\,d^3 k_2\, \delta^{(3)}\left({\bf k}_1+{\bf k}_2-{\bf k}\right)
\,f^\phi_{\rm NL}\left({\bf k}_1,{\bf k}_2\right)
\phi^{(1)}({\bf k}_1)\phi^{(1)}({\bf k}_2)\, ,
\end{equation}
where we have defined 
an effective ``momentum-dependent''
non-linearity parameter $f^\phi_{\rm NL}$.
Here the linear lapse function 
$\phi^{(1)}=\psi^{(1)}$ is a Gaussian random field. 
The gravitational potential bispectrum
reads

\begin{equation}
\langle \phi({\bf k}_1) \phi({\bf k}_2) \phi({\bf k}_3)
\rangle=(2\pi)^3\,\delta^{(3)}\left({\bf k}_1+{\bf k}_2+{\bf k}_3\right)
\,\left[2\,f^\phi_{\rm NL}\left({\bf k}_1,{\bf k}_2\right)\,
{\cal P}_\phi(k_1){\cal P}_\phi(k_2)+{\rm cyclic}\right]\, ,
\end{equation}
where ${\cal P}_\phi(k)$ is the power-spectrum of the gravitational
potential. 

At this point, in order to give the non-linearity 
parameter, an important remark is in order. 
Indeed, when dealing with second-order 
perturbations which are expressed in terms of first-order quantities, 
also the short-wavelength behaviour of the first-order 
perturbations must be taken into account, as it becomes evident when 
going to momentum space. The crucial point here is which is the final 
quantity one is interested in.
We are interested in calculating the bispectrum of 
the gravitational potential on 
large scales as a measure of non-Gaussianity of the 
cosmological perturbations on those scales.  
The bispectrum of such quantities is twice the 
kernel which appears when 
expressing  
the second-order quantities in terms of first-order ones 
in Fourier space, an example of such  a kernel being 
$f^\phi_{\rm NL}\left({\bf k}_1,{\bf k}_2\right)$ in Eq.~(\ref{phimomspace}). 
This means that, when calculating the bispectrum, we can evaluate the kernel 
in the long-wavelength limit, irrespective of the integration over the 
whole range of momenta. This is the reason why the last term in 
Eq.~(\ref{expr:phi2}) gives 
a negligible contribution to the large-scale limit 
of the gravitational potential bispectrum. Therefore,
going to momentum space, from Eq.~(\ref{expr:phi2}) we directly read the 
corresponding non-linearity parameter for scales entering the
horizon during the matter-dominated stage

\begin{equation}
\label{ee}
f^\phi_{\rm NL}\simeq-\frac{1}{2}  +4 
\frac{{\bf k}_1\cdot {\bf k}_2}{k^2}
-3 \frac{\left( {\bf k}_1 \cdot {\bf k}_2 \right)^2}{k^4}+\frac{3}{2} 
\frac{k_1^4+k_2^4}{k^4}
\end{equation}
where $k=\left|{\bf k}_1+ {\bf k}_2\right|$.

The non-Gaussianity provided by expression (\ref{ee}) will add to
the known Newtonian and relativistic 
second-order contributions which are  
relevant  on sub-horizon scales, such as the Rees-Sciama effect
\cite{rs}, whose detailed analysis
has been given  in Refs. \cite{pyne}.

\section{Conclusions}

In this paper we have provided a framework to study the evolution
of non-linearities present in the primordial 
cosmological perturbations seeded by inflation on super-horizon 
scales. The tiny non-Gaussianity generated during the inflationary epoch
driven by a single scalar field 
gets enhanced in the post-inflationary stages giving rise to
a non-negligible signature of non-linearity in the gravitational potentials.
On the other hand, there are many physically motivated inflationary
models which can easily accomodate for a primordial value of $f_{\rm NL}$ 
larger than unity. This is the case, for instance, of a large
class of multi-field inflation models which leads to either 
non-Gaussian isocurvature perturbations \cite{lm} or cross-correlated  
non-Gaussian adiabatic and isocurvature modes \cite{bartolong}. Other 
interesting possibilities include the ``curvaton''
model, where the late time decay of a scalar field other than the inflaton
induces curvature perturbations \cite{curvaton}, and the so-called
``inhomogeneous reheating'' mechanism where the curvature
perturbations are generated by spatial variations of the
inflaton decay rate \cite{zal}.
Our findings indicate that a positive future detection of non-linearity
in the CMB anisotropy pattern will not rule out single field models
as responsible for seeding  structure formation in our Universe.

\vspace*{7mm}
\subsection*{\sc Acknowledgements}
\noindent
It is a pleasure to thank David Wands for several discussions. We also thank 
the referee for her/his  comments which allowed to strengthen the
presentation of our results and to correct for an algebraic error in the 
final formula.
\noindent



\begin{thebibliography}{99}
\bibitem{mw1} 
One can recognize the second-order gauge-invariant curvature perturbation 
(up to a gradient term) recently introduced in Ref. \cite{mw}
\bea
\label{qqq}
\zeta^{(2)}&=&-\psi^{(2)}-\H\frac{\delta^{(2)}\rho}{\rho_0^\prime}
+2\H\frac{\delta^{(1)}\rho^\prime}{\rho_0^\prime}
\frac{\delta^{(1)}\rho}{\rho_0^\prime}\nonumber\\
&+&2\frac{\delta^{(1)}\rho}{\rho_0^\prime}\left(\psi^{(1)\prime}+
2\H\psi^{(1)}\right)-\left(\frac{\delta^{(1)}\rho}{\rho_0^\prime}
\right)^2 \left(\H \frac{\rho_0^{\prime\prime}}{\rho_0^\prime}-
\H^\prime-2\H^2\right)\nonumber\\
&=&-\psi^{(2)}-\H\frac{\delta^{(2)}\rho}{\rho_0^\prime}
-(1+3w)\H^2\left(\frac{\delta\rho^{(1)}}{\rho_0^\prime}\right)^2+4
\H\left(\frac{\delta\rho^{(1)}}{\rho_0^\prime}\right)\psi^{(1)}
\, ,\nonumber
\eea
where in the last passage we have made use of the first-order
continuity equation
(\ref{ll}). The quantity $\zeta^{(2)}$
satisfies the conservation equation $\zeta^{(2)\prime}=4\zeta^{(1)}
\zeta^{(1)\prime}$ leading to $\zeta^{(2)}-2\left(\zeta^{(1)}\right)^2=
$ constant. Since $\zeta^{(1)}$ is conserved on large-scales, this
implies that $\zeta^{(2)}$ is conserved as well. Incidentally, we note that
the combination $\zeta^{(2)}-2\left(\zeta^{(1)}\right)^2$ is equal
to the conserved quantity defined in Ref. \cite{lw}. Of course, 
every quantity differing from $\zeta^{(2)}$ by $c\left(\zeta^{(1)}\right)^2$
with $c$ an arbitrary constant, is constant on super-horizon scales.
\end{thebibliography}
\end{document}